\documentclass[Physsubmission, Phys]{SciPost}

\hypersetup{
    colorlinks,
    linkcolor={red!50!black},
    citecolor={blue!50!black},
    urlcolor={blue!80!black}
}
\usepackage{bm, diagbox, graphicx}
\usepackage[bitstream-charter]{mathdesign}
\DeclareSymbolFont{usualmathcal}{OMS}{cmsy}{m}{n}
\DeclareSymbolFontAlphabet{\mathcal}{usualmathcal}

\begin{document}

\begin{center}{\Large \textbf{
Proton Penetration Efficiency\\
over a High Altitude Observatory in Mexico\\
}}\end{center}

\begin{center}
S. Miyake\textsuperscript{1},
T. Koi\textsuperscript{2},
Y. Muraki\textsuperscript{3$\star$},
Y. Matsubara\textsuperscript{3},
S. Masuda\textsuperscript{3},
P. Miranda\textsuperscript{4},
T. Naito\textsuperscript{5},
E. Ortiz\textsuperscript{6},
A. Oshima\textsuperscript{2},
T. Sakai\textsuperscript{7},
T. Sako\textsuperscript{8},
S. Shibata\textsuperscript{2},
H. Takamaru\textsuperscript{2},
M. Tokumaru\textsuperscript{3} and
J. F. Vald\'{e}s-Galicia\textsuperscript{9}
\end{center}

\begin{center}
{\bf 1} Ibaragi National College of Technology,
Hitachinaka, Ibaraki 312-8508, Japan
\\
{\bf 2} Engineering Science laboratory, Chubu University,
Kasugai, Aichi 487-0027, Japan
\\
{\bf 3} Institute for Space, Earth and Environment, Nagoya University,
Nagoya 464-8601, Japan
\\
{\bf 4} Instituto de Investigaciones Fisicas, UMSA,
La Paz, Bolivia
\\
{\bf 5} Information Science laboratory, Yamanashi Gakuin University,
Kofu 400-8375, Japan
\\
{\bf 6} Escuela Nacional de Ciencias de la Terra, UNAM,
Ciudad Mexico 55010, Mexico
\\
{\bf 7} Physical Science laboratory, Nihon University,
Narashino, Chiba 275-0006, Japan
\\
{\bf 8} Institute for Cosmic Ray Research, The University of Tokyo,
Chiba 277-8582, Japan
\\
{\bf 9} Instituto de Geofisica, UNAM,
04510, Mexico D.F., Mexico
\\
* muraki@isee.nagoya-u.ac.jp
\end{center}

\begin{center}
\today
\end{center}


\definecolor{palegray}{gray}{0.95}
\begin{center}
\colorbox{palegray}{
  \begin{tabular}{rr}
  \begin{minipage}{0.1\textwidth}
    \includegraphics[bb=0 0 1000 500, width=30mm]{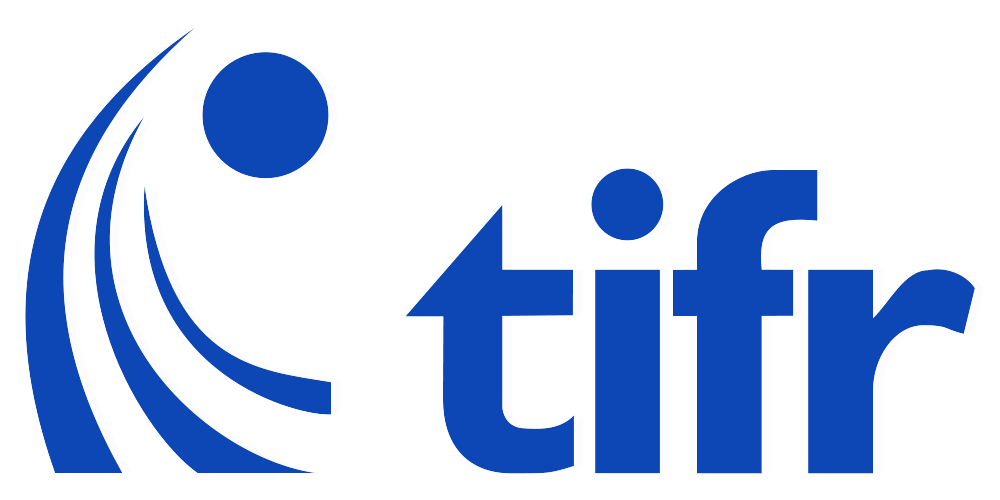}
  \end{minipage}
  &
  \begin{minipage}{0.85\textwidth}
    \begin{center}
    {\it 21st International Symposium on Very High Energy Cosmic Ray Interactions (ISVHE- CRI 2022)}\\
    {\it Online, 23-27 May 2022} \\
    \doi{10.21468/SciPostPhysProc.?}\\
    \end{center}
  \end{minipage}
\end{tabular}
}
\end{center}

\section*{Abstract}
{\bf
In association with a large solar flare on November 7, 2004,
the solar neutron detectors located
at Mt.~Chacaltaya (5,250\,m) in Bolivia
and Mt.~Sierra~Negra (4,600\,m) in Mexico
recorded very interesting events.
In order to explain these events,
we have performed a calculation solving the equation of motion
of anti-protons inside the magnetosphere.
Based on these results,
the Mt.~Chacaltaya event may be explained by the detection of solar neutrons,
while the Mt.~Sierra~Negra event may be explained by the first detection
of very high energy solar neutron decay protons (SNDPs) around 6\,GeV.
}

\section{Introduction}
\label{sec:intro}
An interesting event was registered in association with the large solar flare
on November 7, 2004 by the high altitude solar neutron detectors
located at Mt.~Chacaltaya in Bolivia and Mt.~Sierra~Negra in Mexico
\cite{Muraki2021},\cite{ Muraki2020}.
The data are shown in {bf Figure 1} and {bf Figure 2}
respectively.
Counting rate excesses in both detectors started
at the same time around 15:50\,UT,
however clear differences were observed
in the duration of the respective events.
The Chacaltaya event lasted for 20\,minutes,
while the Sierra~Negra event continued for 78\,minutes.
The signal of the Chacaltaya event may be explained
by the detection of solar neutrons.
These neutrons were produced at 15:47\,UT on the solar surface
instantaneously with the increase of X-ray intensity.

\begin{figure}[ht]
\centering
\includegraphics[width=110mm]{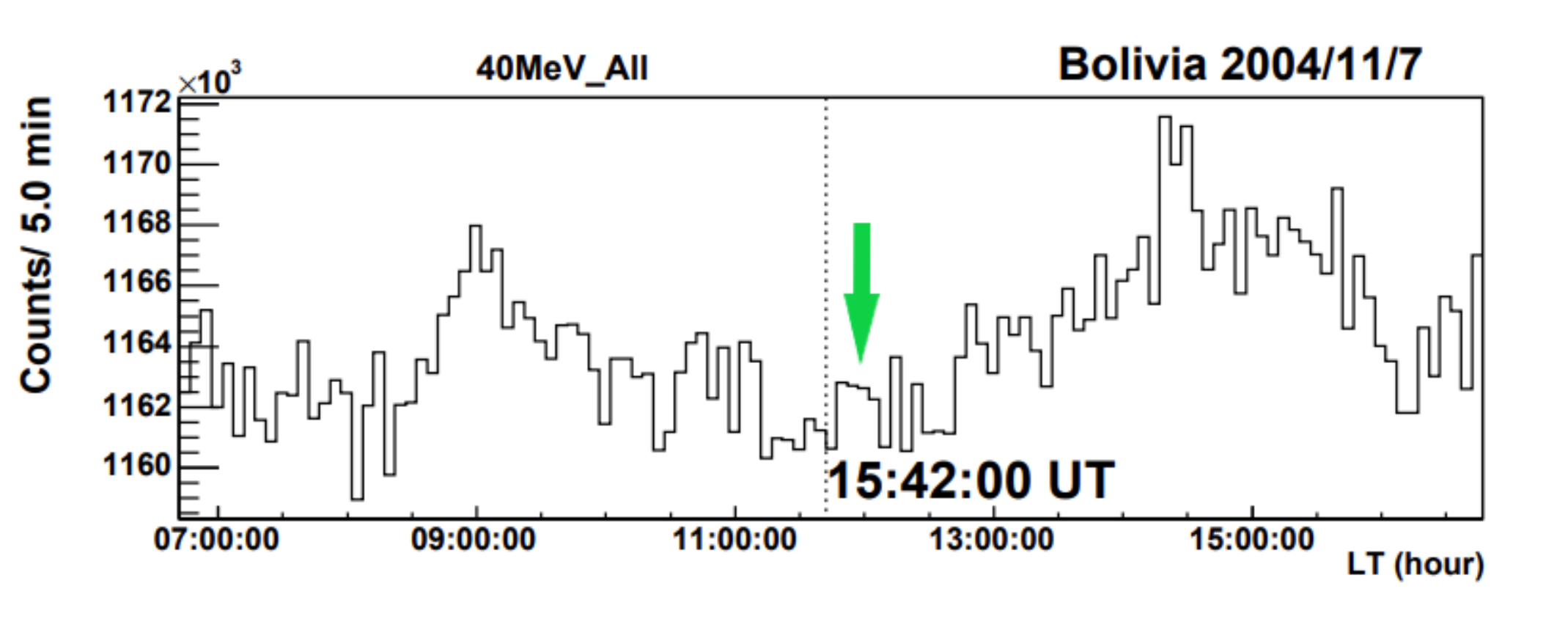}
\caption{%
The 5-minute value of the counting rate of the solar neutron detector
located at Mt.~Chacaltaya (5,250\,m).
The first peak at the local time 9\,am
corresponds to the particle current along the IMF line,
the second peak at 12\,LT (the green arrow) was produced by solar neutrons,
and the third peak around 14:30\,LT corresponds
to the arrival of the CME to the Earth.
The threshold of this channel corresponds to the particles
with energy higher than 40\,MeV.
The flare start time (15:42\,UT) is shown by the dotted line.}
\label{fig-1}
\end{figure}

\begin{figure}[ht]
\centering
\includegraphics[width=110mm]{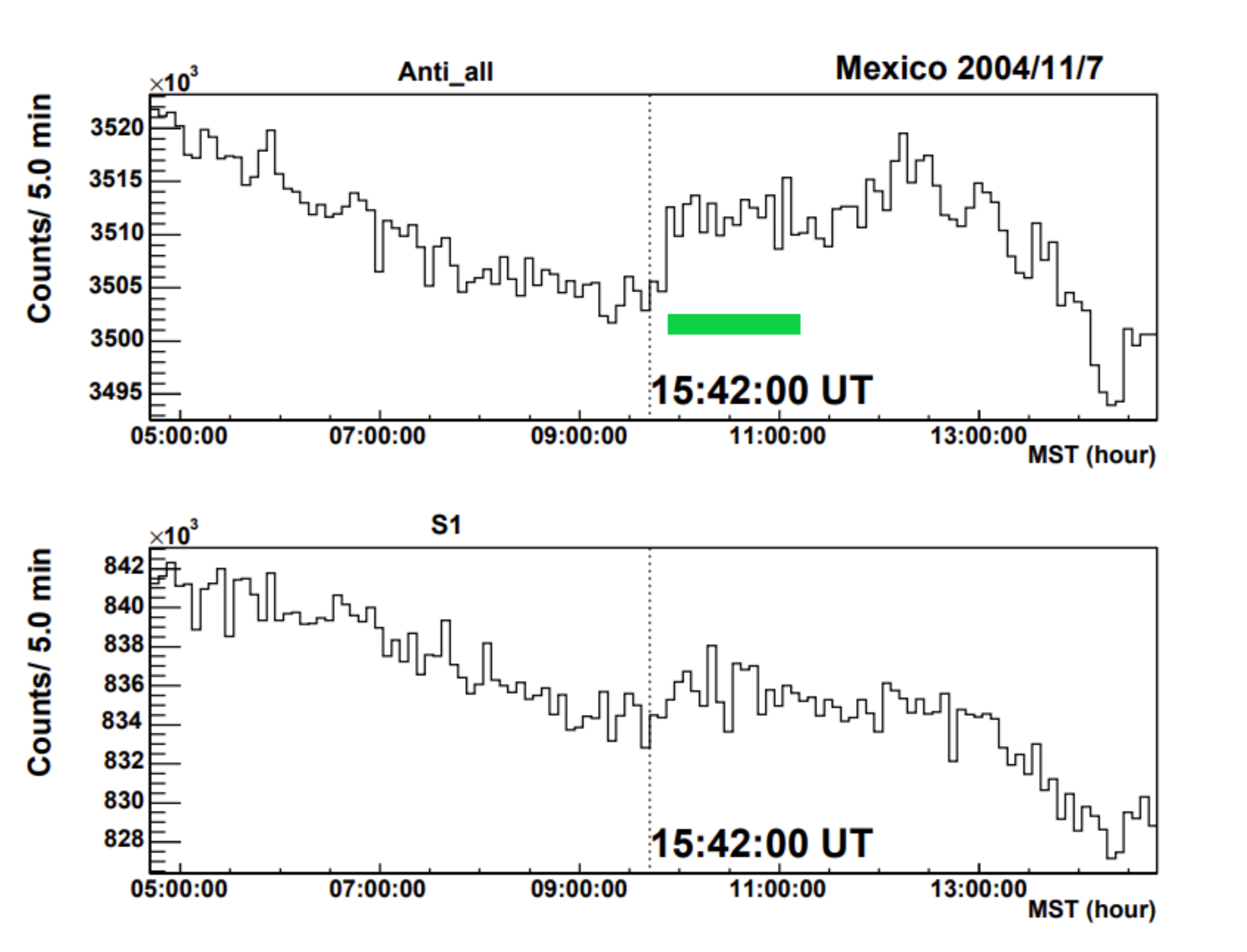}
\caption{%
The 5-minute value of the anti-counter and the lowest proton-neutron channel (S1)
of the Mt.~Sierra~Negra solar neutron telescope on November 7th of 2004.
The horizontal green line indicates the time span
when the excess was observed ($\sim$\,78\,minutes).}
\label{fig-2}
\end{figure}

If the Sierra~Negra event was produced by solar neutrons,
the excess should not continue after 25\,minutes,
since the threshold energy of one of the channels of the detector (S1)
was set at $>$\,30\,MeV.
Therefore, we have assumed
that the excess of the Sierra~Negra detector may be explained
by the detection of Solar Neutron Decay Protons
(SNDPs) \cite{Evenson1983, Droge1996}.

If the observed excess counts are really produced by protons,
we must show how they can arrive at Mt.~Sierra~Negra,
passing through the magnetosphere.
The cutoff rigidity of the magnetic latitude of Mexico
was originally calculated as 8\,GV \cite{Stormer1955}.
However, an early work by Smart, Shea, and Fl\"{u}ckiger \cite{Smart2000}
suggests a possibility that low energy protons less than
the rigidity of 8\,GV
could penetrate into the magnetosphere
and arrive over the atmosphere of Mt.~Sierra~Negra
\cite{Cardenas2012, Gonzalez2015}.
Therefore, we estimate the detection efficiency
of low energy protons in the energy range between 4.5\,GeV and 20\,GeV.

In the next section,
we describe details of the calculation and present the results.
Then we compare the results of the calculation
with the two experimental results. 
We examine whether both events are reasonably explained
by the hypothesis of Solar Neutron Decay Protons.

\section{Calculation Method and Results}
\label{sec-2}

\noindent
{\it Method}:
We have injected anti-protons from 20\,km above Mt.~Sierra~Negra.
Anti-protons were emitted every one degree
in the north-south direction
and the east-west direction independently.
Therefore for one fixed energy of anti-protons,
32,761 ($181 \times 181$) trajectories were examined.
The trajectory of each anti-proton was followed
by solving the equation of motion using the Runge-Kutta-Gill method
until they arrive at the magnetopause at $8\,R_{\rm E}$
({\it allowed}) \cite{Miyake2017}.

Of course some trajectories do not reach at $8\,R_{\rm E}$.
Then they were counted as the forbidden trajectories for proton arrivals.
The initial energy of anti-protons was examined in the energy range
between 4.5\,GeV and 20\,GeV.
In the present calculation,
the distance from the Earth center to the head of the magnetosphere
({\it i.e.} magnetopause) is approximated by $8\,R_{\rm E}$
and examined whether or not anti-protons arrived there.
The SNDPs are expected to come from the day side,
so present approximation may be enough for this study.

\begin{figure}[ht]
\centering
\includegraphics[width=80mm]{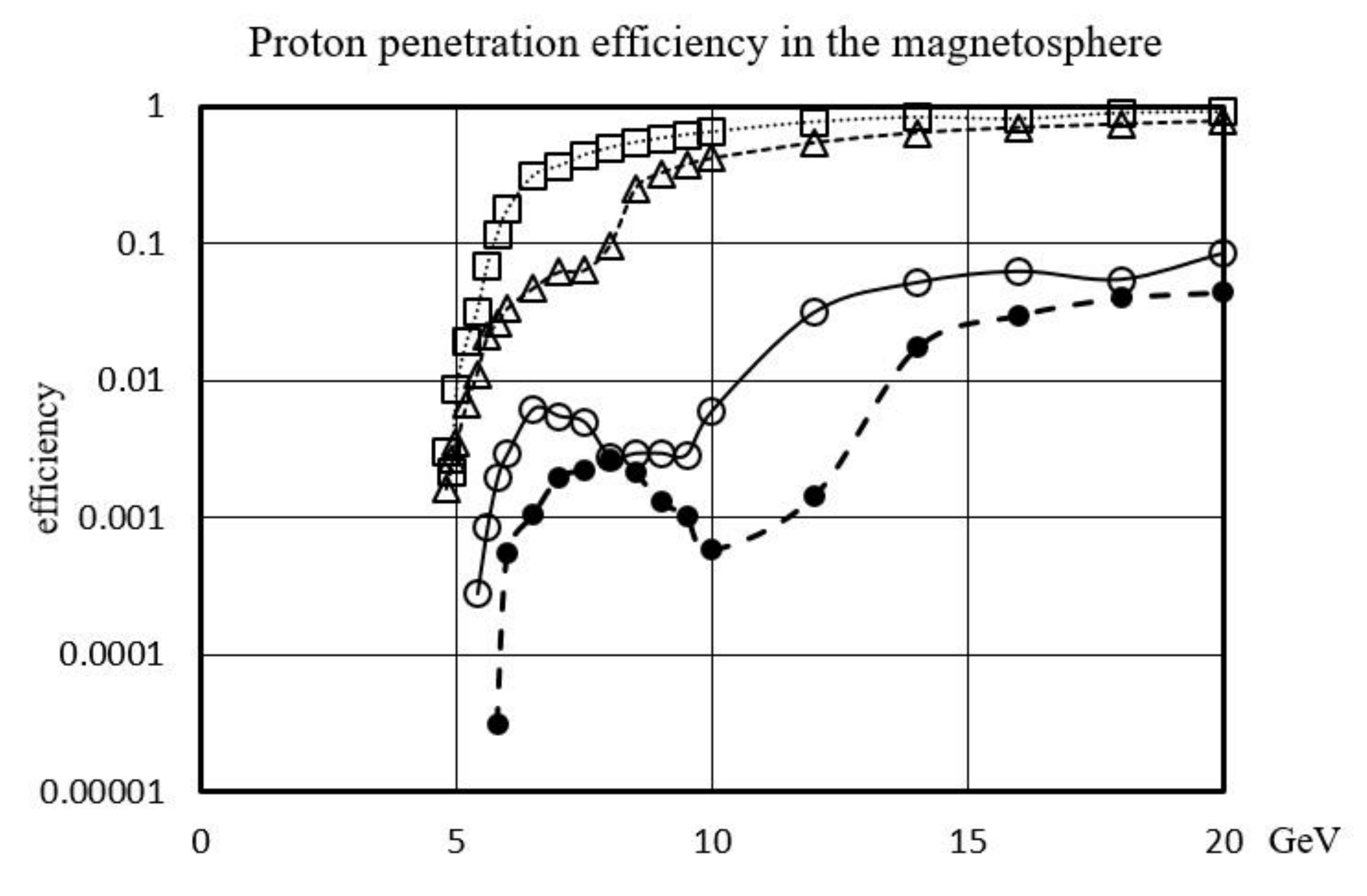}
\caption{%
Proton penetration rate is shown as a function of the incident energy.
The open box presents the protons coming into the top of the atmosphere
of Mt.~Sierra~Negra from all directions.
The open triangles represent protons arriving from the day side.
These data are normalized by the shot number, 32,761 examples.
On the other hand, the proton rate entering over the top of the atmosphere
with the incident angle less than 20 degrees is shown by closed circles.
The protons rate coming at the top of the atmosphere
with an incident angle between 20-40 degrees are shown
by open circles.}
\label{fig-3}
\end{figure}

\noindent
{\it Results}:
We found that quite low energy anti-protons,
less than 8\,GeV, arrived at the magnetopause
as predicted by the earlier work \cite{Smart2000}.
The proton penetration probability from all directions
at 20\,km above Mt.~Sierra~Negra is presented by open boxes
in {\bf Figure~\ref{fig-3}},
while the arrival probability from the day side ($X > 0$)
is given by open triangles. 

\begin{figure}[ht]
\centering
\includegraphics[width=80mm]{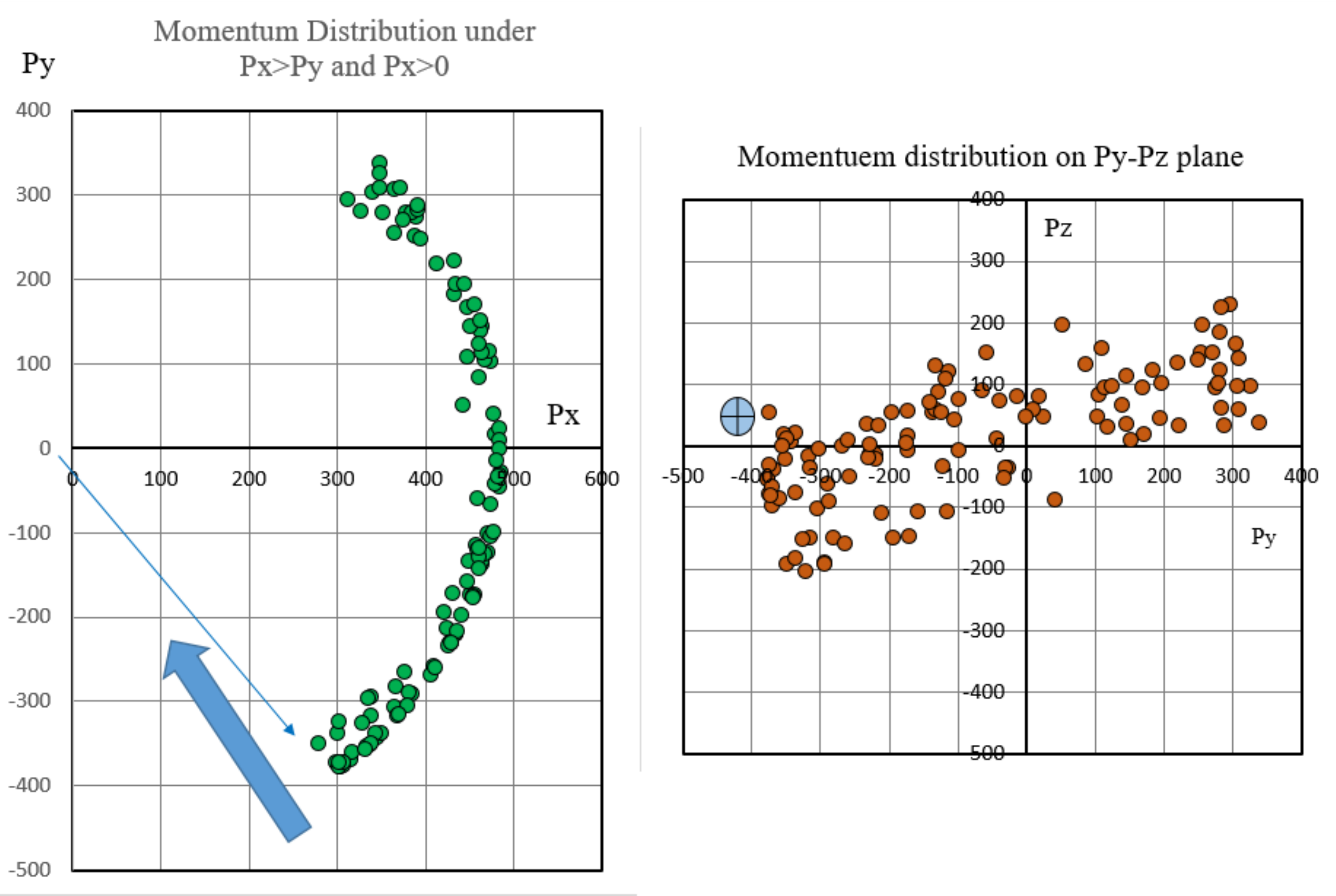}
\caption{%
The momentum distribution of anti-protons arrived at 8$R_{E}$
is shown on $P_{X}-P_{Y}$ plane (left) and $P_{Z}-P_{Y}$ plane (right) respectively.
Here the positive $P_{Y}$ is defined to the eastward,
while $P_{Z}$ positive vector points the northward.
The $P_{X}$ axis indicates the solar direction.
The IMF direction (${56}^{\circ}\sim{60}^{\circ}$) is shown by the arrow
and by the $\oplus$ mark in the plots respectively.}
\label{fig-4}
\end{figure}

We also take into account the crossing angle
between $X$-axis of GSE coordinate and the momentum (${\bf P}$) of SNDPs.
Taking into account the entrance of charged particles
along the IMF direction (${45}^{\circ}\sim{60}^{\circ}$),
the anti-protons to satisfy the two conditions
of $P_{X}>0$ and $\tan^{-1}{(P_{X}/P_{Y})}$ larger than ${45}^{\circ}$
were finally selected.
(In other words, the momentum region of $P_{Y}<P_{X}$ is selected.)
The results are shown in {\bf Figure~\ref{fig-4}}
on the $P_{X}-P_{Y}$ plane and $P_{Y}-P_{Z}$ plane
of the GSE coordinate respectively.
We require further condition;
the incident angle to the atmosphere of the incident protons
is less than $40^{\circ}$.
All points plotted in Figure~\ref{fig-4} satisfy these conditions.

Furthermore, we take into account another factor;
proton attenuation in the atmosphere.
When protons enter into the air vertically,
the survival probability of proton signal is larger
than the arrivals from large zenith angles.
The value is estimated as to be approximately 0.4 for neutrons
with vertical entrance ($\theta < 20^{\circ}$)
and 0.2 for the entrance with $40^{\circ}$ respectively.
Those results were obtained by using the GEANT4 simulation.
Details are given in Supplemmentary Information (S1) [arXiv].

\section{Boosting factor and Reduction factor}
\label{sec-3}

\begin{figure}[ht]
\centering
\includegraphics[width=80mm]{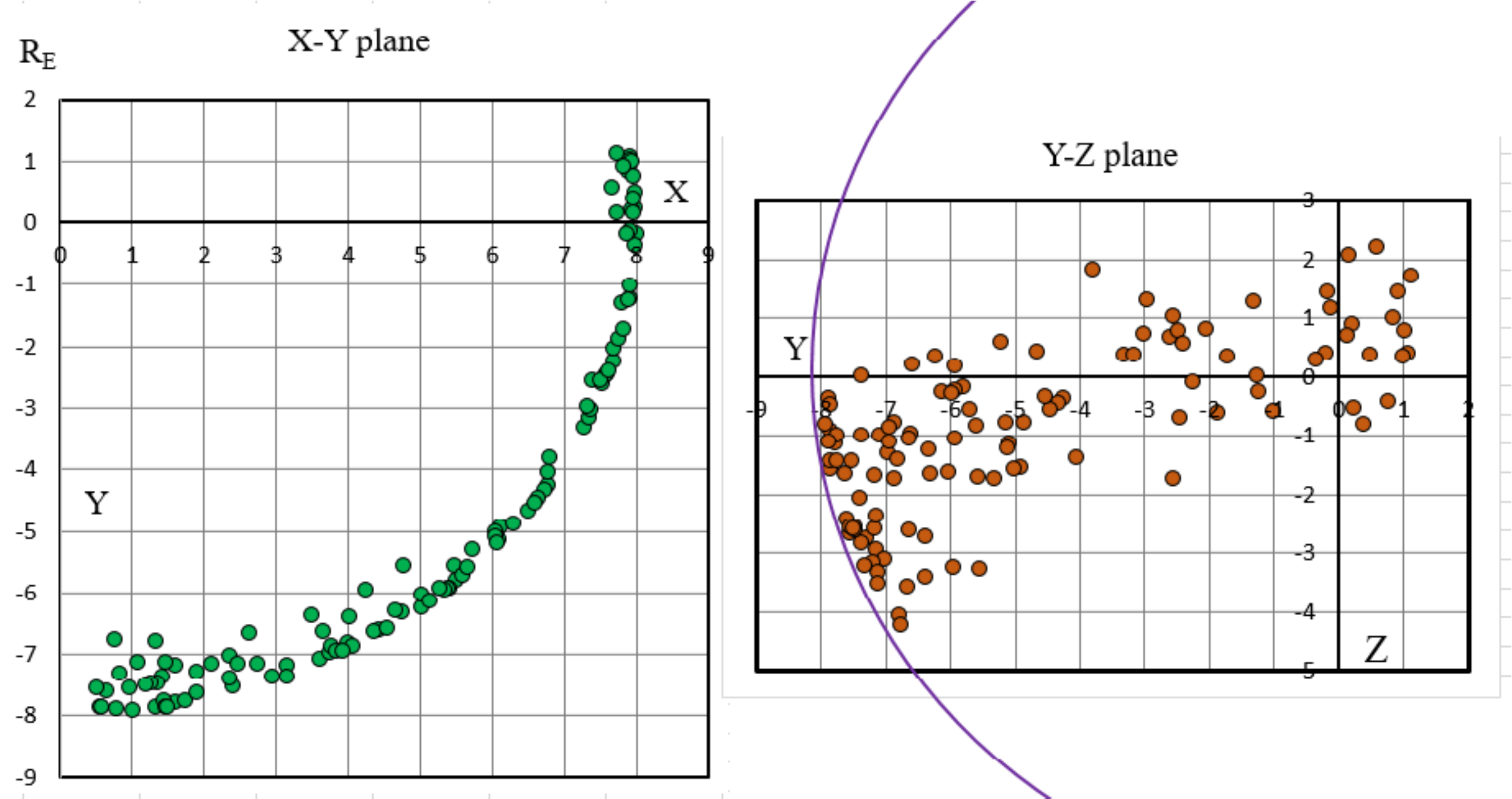}
\caption{The arrival map of anti-protons of 6\,GeV
at 8\,$R_{\rm E}$ on the $X-$Y plane,
while the right side plot shows the distribution on the $Y-Z$ plane.
The $Z$ axis corresponds to the north-south direction,
while $X$-axis directs toward the Sun.
Here the equivalent incident angle of protons to the atmosphere
is selected less than 40 degrees.
The number of the data points is 126,
however when we require a condition
of Vy $<$ 0, this number is reduced to 46.}
\label{fig-5}
\end{figure}

\noindent
{\it Boosting factor}:
Figure~\ref{fig-5} presents the arrival point of the anti-protons
at 8\,$R_{E}$.
Figure~\ref{fig-5} tells us another information.
The acceptance area of the interplanetary protons by the magnetopause
does not cover all area of the magnetopause (not circle as in Figure 5)
but the acceptable area is limited within the rectangular area.
Thus total area of acceptance of the magnetopause for receiving the SNDPs
is estimated as to be $1\times{10}^{15}\,{\rm m^2}$.
The decay factor of neutrons with the energy of 6\,GeV
during the flight in the distance of ${\it l}\sim{\rm 0.067\,au}$,
is estimated as 0.0047. More details are given in
Supplementary Information 2 and 3 (S2, S3) [arXiv].

After we apply the decay factor to the above area,
the effective decay area for accepting the SNDP signals around 6\,GeV
may be evaluated as to be $4.7\times{10}^{12}\,{\rm m^2}$.
The detector areas at Mt.~Chacaltaya
and Mt.~Sierra~Negra are only $4\,{\rm m^2}$.
Therefore, in comparison with these small detector areas,
a huge collecting area will be expected according to our estimation
that may intensify very weak signal
of high energy neutrons.
Here let us call the effect as a {\it Horn effect}.
Details are also described in (S3) [arXiv].

\noindent
{\it Reduction factor}:
Assume that one high energy neutron decay proton
is produced in a unit long volume in the space
(1\,SNDP/($\ell\cdot{\rm m^2}$)),
then enormous amount of protons will be produced.
(Let us imagine a cylinder space with the base of $1\,{\rm m^2}$.)
According to the above estimate,
the number of protons must be of the order of $4.7\times{10}^{12}$.
But not all of them can enter into the magnetosphere.
The Earth has a capability to protect ``cosmic radiation''
through the double gates;
(a) by the absorption in the air and
(b) by the rejection with the magnetic field.
By the former process of (a), the flux of incident protons will be reduced
by an order of $0.1\,\sim\,0.4$ depending on 
the incident angle to the atmosphere (see S1),
while by the latter process of (b),
incoming protons will be rejected by the condition of,
{\it i.e.}, the incoming proton trajectory
must be continiously connected with the trajectory inside the magnetosphere.

As is shown in S1, the reduction factor by the absorption in the air
depends on the incident angle.
In case protons enter within $40^{\circ}$,
the candidates of the well connected trajectory may be reduced
to 46 among the 32,761 simulated trajectories
as shown in Figures~\ref{fig-4} and \ref{fig-5}.
Thus, the entrance probability will be reduced
to be less than $1.4\times{10}^{-3}$.
If we require the entrance condition of protons into the air
less than $20^{\circ}$, only 10 events are left.

The momentum vector of the anti-protons must match with the momentum vector
of protons arriving from the outside of the magnetosphere.
As shown in Figure~\ref{fig-4},
the matching probability of the two tracks is very few and actually
no optimum vector is found in
the $P_{X}-P_{Y}$ plane of Figure~\ref{fig-4} that matches with the IMF direction.
Therefore, we postultate here
the number of matching trajectory is less than one.
Then actual rejection factor by above condition
reduces the acceptable flux to be less than $3.1\times{10}^{-5}$ (= 1/32,761).
We should collect more samples of the simulation
around the allowed condition region to get a finite number.

In addition, we require another condition of the smooth connection
to the momentume vector in the  $P_{Z}-P_{Y}$ plane,
Because protons will make gyro-motion along the IMF direction.
The momentum vector of the $Z$-direction
varies to the north-south direction.
Details are given in Supplementary Information (S4) [arXiv].
As a result, the total reduction factor can be estimated as to be
$9.6\times{10}^{-10}$
($= 3.1\times{10}^{-5}\times 3.1\times{10}^{-5}\times 1.4{10}^{-3}$).
Therefore when we multiply this reduction factor to the boositing fcator,
we may get the ``effective'' boosting factor of SNDPs as to be 450
($= 9.6\times{10}^{-10}\times 4.7\times{10}^{12}$).

In summary,
the detection efficiency of SNDPs can be described
by the production of the two factors;
the entrance probability of the SNDPs into the magnetosphere
from the interplanetary space
{\it by} the attenuation of the SNDPs inside the atmosphere.

\section{Application of Results to Actual Data}
\label{sec-4}

Let us compare our prediction with the observed results.
In this chapter,
we examine whether current estimation may explain the observed results.

From the observed data of Mt.~Chacaltaya,
the neutron intensity is estimated as
$1.5\times{10}^{6}/{\rm m^{2}}$ at 100\,MeV
and $3,000/{\rm m^{2}}$ at 1,000\,MeV (1\,GeV) respectively.
These intensities were already converted into the flux
at the top of the atmosphere,
taking into account the attenuation in the atmosphere.
However the Chacaltaya detector did not measure the high energy region
beyond 1\,GeV.
Therefore, we estimate the flux at 6\,GeV
by extending the Chacaltaya spectrum into high energies.
We estimate the flux to the three cases of the spectra beyond 1\,GeV,
assuming the integral spectrum of the power law as $E_{n}^{-\gamma}$,
with the power index of $\gamma$.
Case (1) simple extension from 1 to 10\,GeV with $\gamma=2.7$,
Case (2) moderate case; $\gamma=3.7$,
and Case (3) Soft case; $\gamma=4.7$.
Then we may estimate the probable flux at 6\,GeV for
Case (1) = $23.7/{\rm m^{2}}$,
Case (2) = $4.0/{\rm m^{2}}$,
and Case (3) = $0.66/{\rm m^{2}}$ respectively.
See also Supplementary Information (S5) [arXiv].

Now we compare the extended flux of neutrons at 6\,GeV
with the flux of the SNDPs observed at Mt.~Sierra~Negra.
From the actual data of the S1 channel,
the total flux of SNDPs may be estimated as $(45,000 \pm 900)/{\rm m^{2}}$
or the first one-minute value may be deduced
as $(360 \pm 125)/({\rm m^{2}\cdot min.})$.
The latter flux may correspond to the observed events in early time.
They were the decay product of neutrons with the energy greater than 6\,GeV
between 0.934\,au and 1.0\,au ($\approx$ 0.066\,au).

Then we find that the observed one-minute value of Mt.~Sierra~Negra
is 14$\sim$\,1,130 times higher
than the extended flux of Chacaltaya.
Taking into account present calculation of the boosting factor
intensified by $\sim$\,450 times,
the observed results could fairly well reproduce the experimental result.
The expected boosting factor can explain actual observed data.

\section*{Acknowledgements}
The authors acknowledge the staffs of Mt.~Chacaltaya cosmic ray observatory
and Mt.~Sierra~Negra observatory
for keeping the detectors in good condition.
This event was re-analyzed,
being inspired after a lecture
by Prof. Sunil Gupta and Dr. Pravata Mohanty
when they visited Nagoya University in February 2020.


\section{Supplementary Information}
\label{sec-6}

\noindent
{\it Supplementary Information 1}
 
Incident protons make nuclear interactions
at the top of the atmosphere and lose the energy.
As a result, the arrival signals are lost and not received
by the detectors located at the ground level.
We have examined the attenuation rate of a few GeV proton
in the atmosphere by the GEANT4 simulation code.

\begin{figure}[ht]
\centering
\includegraphics[width=110mm]{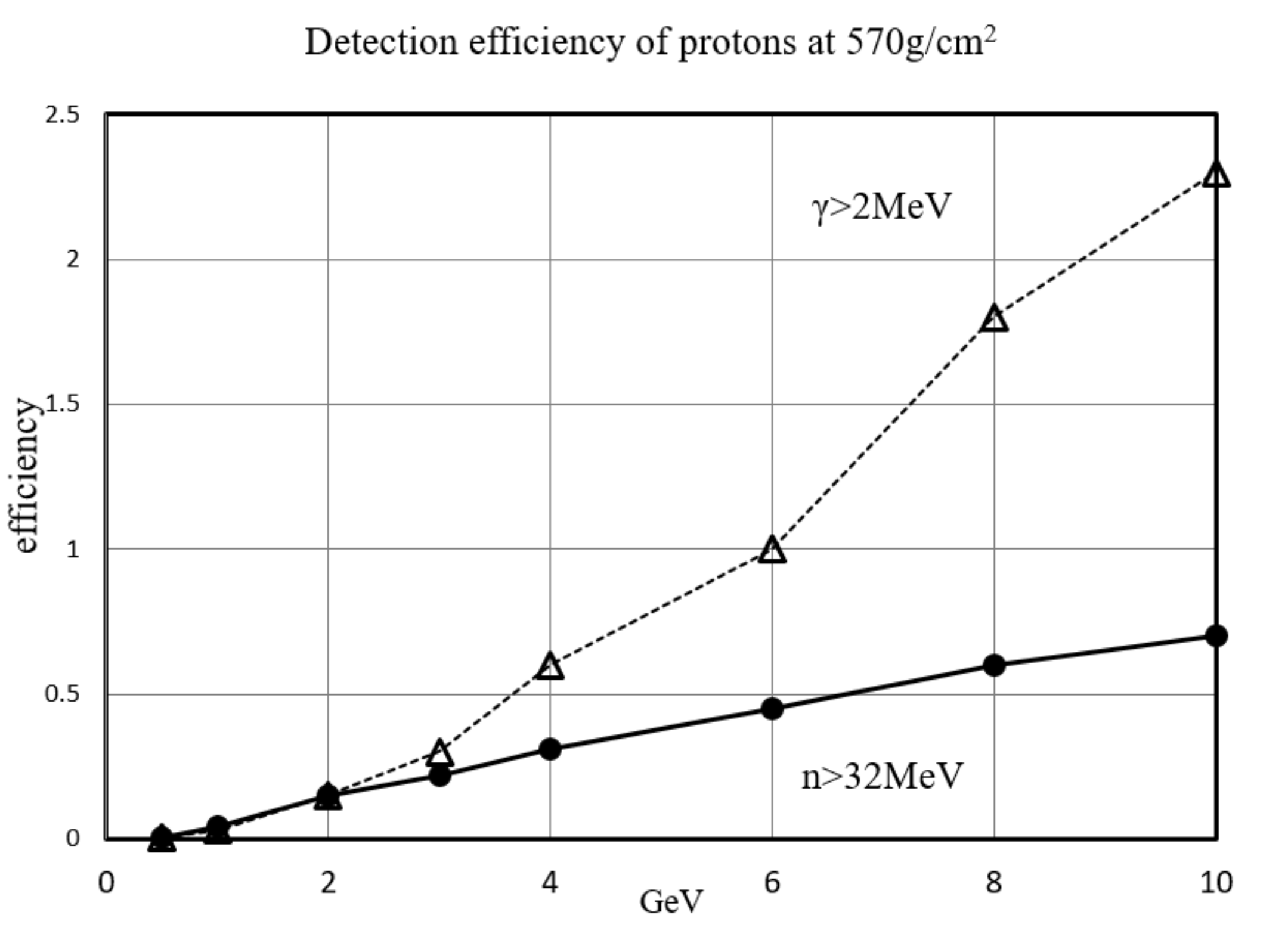}
\caption{%
The detection efficiency of
low energy gamma rays ({\large $\bigtriangleup$})
and neutrons ({\large $\bullet$})
by the Sierra~Negra solar neutron telescope
for vertical entrance into the air is shown
as a function of the incident energy of protons.
The discrimination level of gamma-rays and neutrons
is set at larger than 2\,MeV and 32\,MeV respectively.}
\label{sf1}
\end{figure}

\begin{figure}[ht]
\centering
\includegraphics[width=110mm]{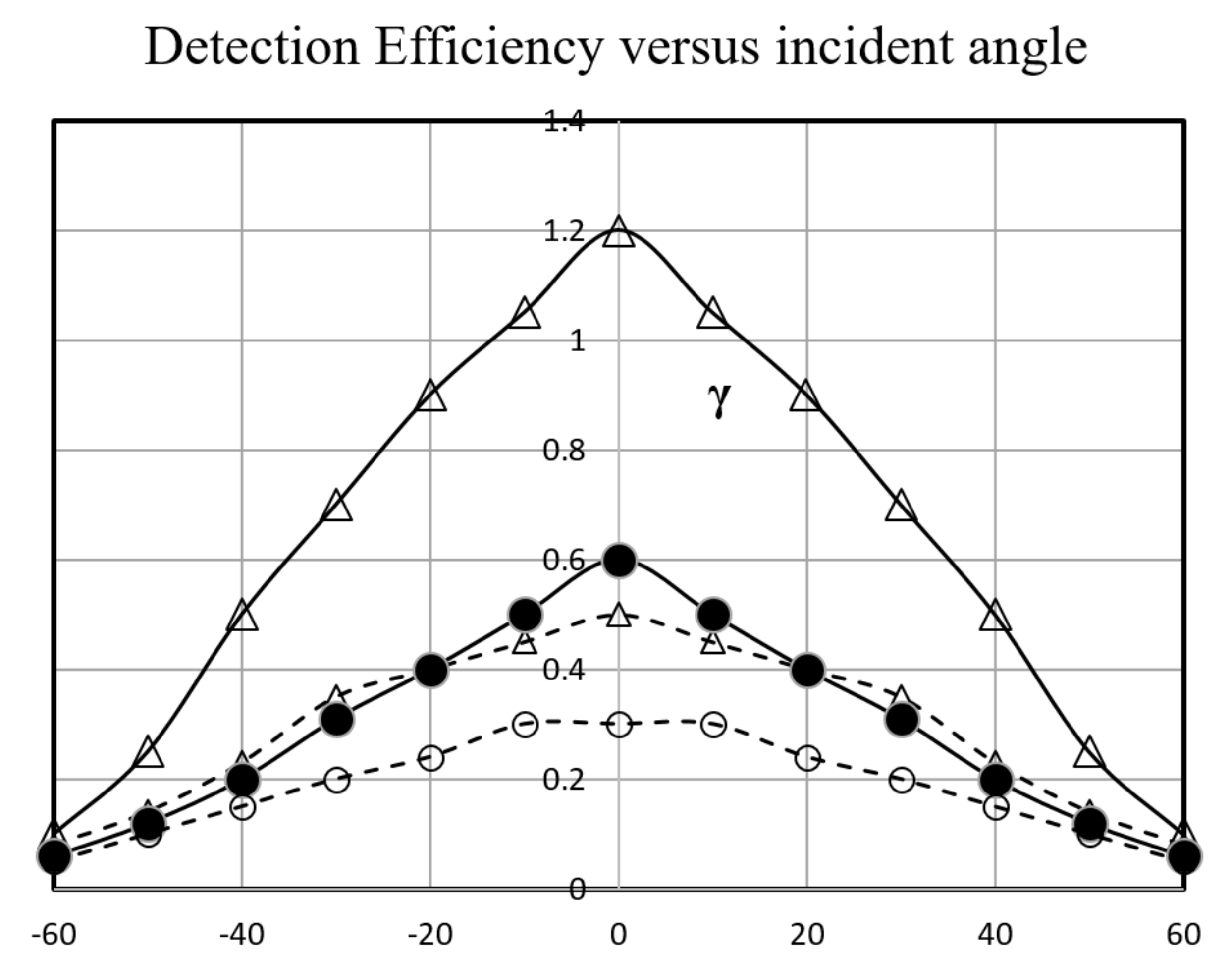}
\caption{%
The detection efficiency of incident protons is shown
as a function of the incident angle
on the top of the atmosphere.
Open circle and closed circle correspond to neutrons
with the energy higher than 32\,MeV
for the incident energy 4\,GeV ({\large $\circ$})
and 6\,GeV ({\large $\bullet$}) respectively,
while the triangles correspond to the gamma-rays
with the energy higher than 2\,MeV
for the incident energy of protons of 4\,GeV ({\large $\triangle$})
and 6\,GeV ({\large $\bullet$}) respectively.}
\label{sf2}
\end{figure}

The results are given in Figure~\ref{sf1} and Figure~\ref{sf2}.
The Figure~\ref{sf1} presents the detection efficiency of signal
of low energy protons at 4,600m altitude.
In the Figure~\ref{sf1},
we assumed that protons enter into the atmosphere vertically
($\theta\approx\,{0}^{\circ}$)
and the detection is made either by using the ``anti-all'' channel
or by the ``channel 1''.
In the anti-all channel,
shower debris like soft gamma-rays are detected by PR-counter,
while in the channel ch-1,
neutral particles with energy higher than 30\,MeV
are detected by the thick plastic scintillator.
For 6\,GeV protons by the ch-1,
the detection efficiency is predicted as about 0.4
for the vertical entrance.

The attenuation rate strongly depends on the incident angle.
Therefore we have calculated the attenuation of the proton signal
as a function of the incident angle.
Figure~\ref{sf2} shows the angular dependence
of $E_{p}=4$\,GeV and 6\,GeV protons.
In the case, protons enter with the zenith angle of 40 degrees,
only 10 percent of signals are recorded
by the ground based detector located at Mt.~Sierra~Negra.

\noindent
{\it Supplementary Information 2}

\begin{figure}[ht]
\centering
\includegraphics[width=110mm]{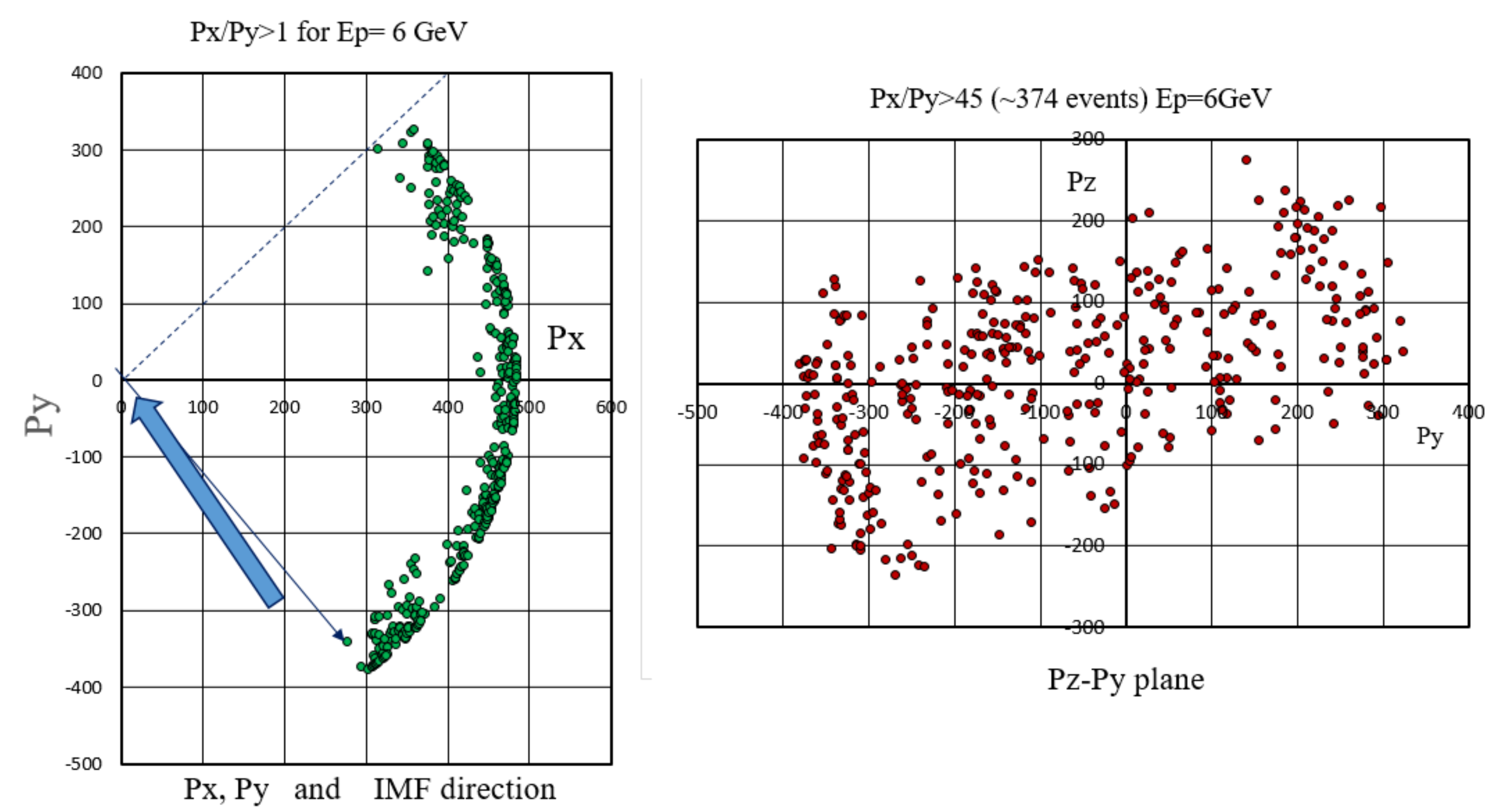}
\caption{%
(left side)
Among 32,761 shots, 374 events were selected
to satisfy the condition
of $\tan^{-1}{(P_{X}/P_{Y})} > 45$ degrees and $P_{X} > 0$
for $E_{p}={\rm 6\,GeV}$ anti-protons.
They are plotted on the $P_{Z}-P_{Y}$ plane.
(right side)
The same events are plotted on the $P_{Z}-P_{Y}$ momentum space.
No incident angular cut of the incoming protons to the air was applied.
The data involve all arrival directions at 20\,km above Mt.~Sierra~Negra.
The IMF direction is shown by the arrow.}
\label{sf3}
\end{figure}

Figure~\ref{sf3}
shows the momentum distribution of anti-protons arrived at $6\,R_{E}$.
In this plot,
we do not require any condition on the incident angle,
so that in the plot all anti-protons are included.
They were injected even almost horizontally from 20\,km above Mt.~Sierra~Negra.

\begin{figure}[ht]
\centering
\includegraphics[width=110mm]{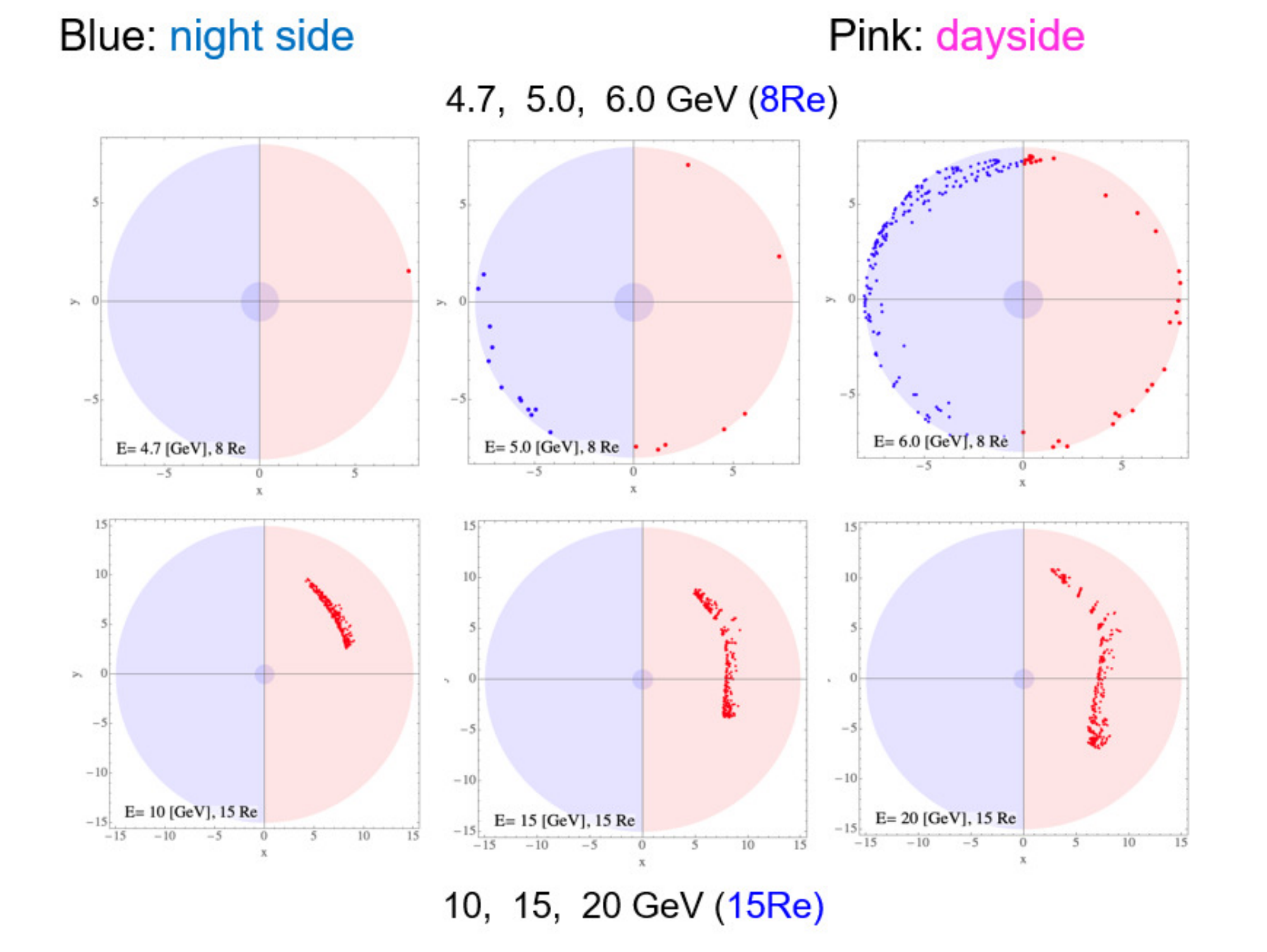}
\caption{%
The outgoing points of anti-protons at $8\,R_{\rm E}$ (radius of the Earth)
and $15\,R_{\rm E}$ at the magnetopause are shown by dots.
The right side corresponds to the day side of the Earth,
while left side corresponds to the night side.
In the low energy region,
a number of protons are coming from the night side.
On the other hand, the incident momentum increases,
they are arriving to the Earth only from the day side.}
\label{sf4}
\end{figure}

Figure~\ref{sf4}
presents the arrival map of anti-protons at $8\,R_{E}$
with the energy of 4.7,\,5,\,6\,GeV respectively.
Quite a lot of anti-protons arrived to the night region
of the Earth (the blue side) in the low energy,
while the proton energy increases from 10,\,15,\,to 20\,GeV,
the arrival direction concentrates into the day side (the pink area).

\noindent
{\it Supplementary Information 3}

In this event, it is quite important to understand general situation
of the solar terrestrial environment.
Just one day before of X2 solar flare,
M9 class solar flare occurred at the Sun.
Following this M9 flare, a large CME was emitted.
Here we call the CME as CME1.
The CME1 approaches up to $\ell\approx\,10^{7}\,{\rm km}$
in front of the Earth.
The strength of the magnetic field was measured
by the magnetometer of the ACE satellite \cite{ACE}
and the GEOTAIL satellite \cite{Geotail}.
The measured field strength of the CME rope was 50\,nT,
while the field strength near the Earth
between 16:00-18:30\,UT was measured as 20\,nT.

The decay probability of neutrons with $E_{n}=6$\,GeV is estimated as 0.0047.
The probability is very small.
If we put the field strength of $H={\rm 20\,nT}$
and proton energy of $p=6$\,GeV in the formula
of $p_{\perp} [{\rm GeV}]=0.3 H [{\rm T}] \rho [{\rm m}]$,
the rotation radius of protons is $1\times\,10^{6}$\,km.
In comparison with $\ell$, it is about 10 times shorter.
Therefore, until they arrived at the magnetopause,
the protons rotated about $\sim$\,8 times as the maximum case.

We also know the field strength in the CME as 50\,nT.
Protons less than 60\,GeV may be trapped in the CME rope.
For this reason, the SNDPs that appeared between the Sun and the CME1
could not come to the Earth.
This scenario leads us to establish the decay length of neutrons
as $\ell\approx\,10^{7}\,{\rm km}$ or 0.067\,au.

\begin{figure}[ht]
\centering
\includegraphics[width=110mm]{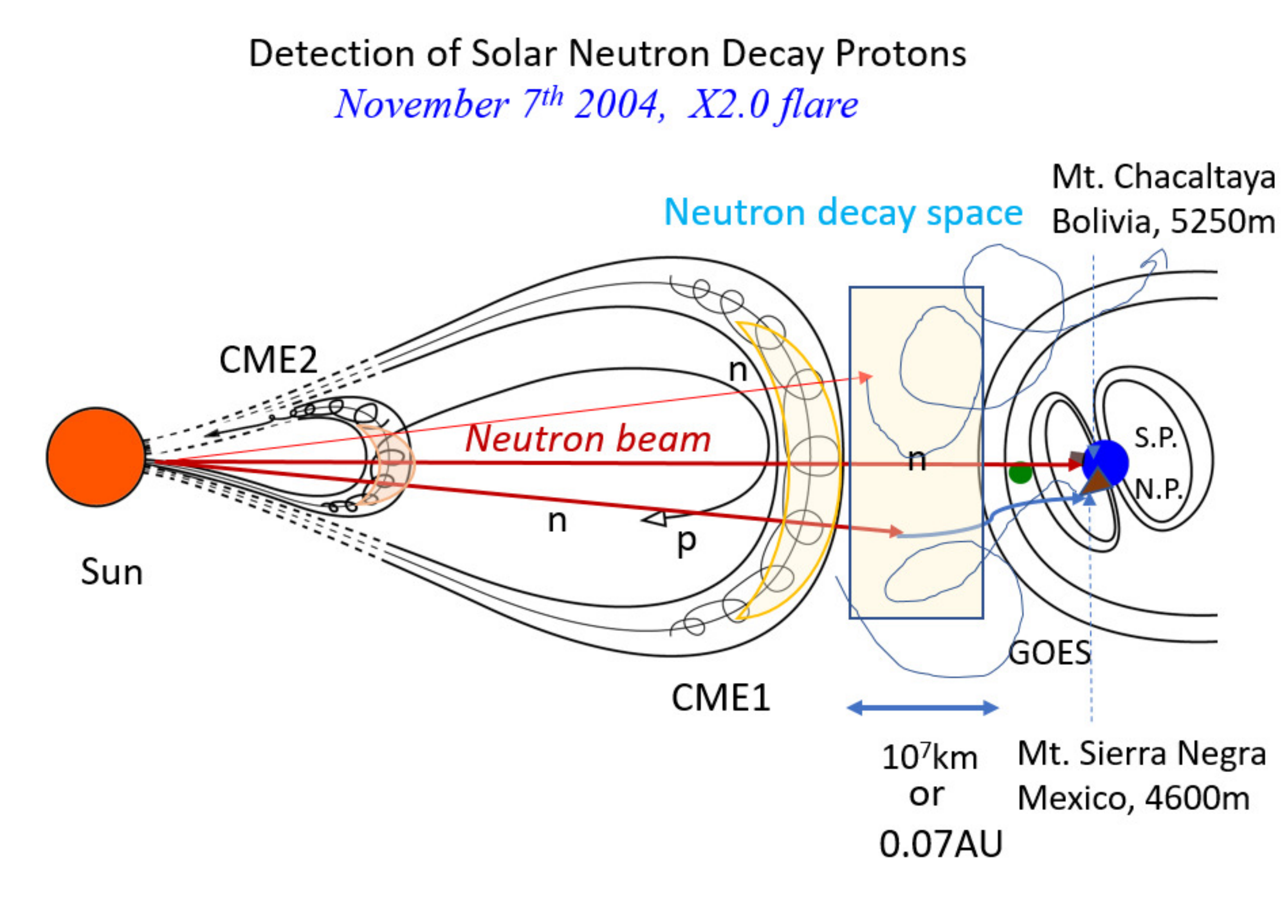}
\caption{%
The interplanetary space around 16\,UT on November 7th, 2004
is pictorially presented.
The CME1 was produced one day before in association with a M9.7 flare.
On the other hand, the CME2 was emitted on 16\,UT of November 7th 2004,
by the X2.0 flare.
The neutron beam penetrates these plasma wave,
however protons, the decay products of solar neutrons,
cannot penetrate the {\it CME wall} (CME1)
due to its strong magnetic field.
Therefore, the decay length is estimated as $1\times10^{7}km$.
The effective decay space is also shown by the blue box.}
\label{sf5}
\end{figure}

In comparison with low energy protons
produced by the solar neutron decay process,
the behavior is quite different
from previous low energy events \cite{Evenson1983, Droge1996}.
Therefore we list  the numerical values
to estimate the difference in Table~\ref{tab-1}.
The situation is summarized and depicted
in Figure~\ref{sf5}.

\begin{table}[ht]
\centering
\caption{The gyro-radius of protons in the magnetic field
at the magnetopause and magnetosheath}
\label{tab-1}
\begin{tabular}{ccccc}
\hline
$B\quad |\quad E_{\rm p}$& 10\,MeV & 100\,MeV & 1\,GeV & 6\,GeV \\
\hline\hline
{\ }{\ }5\,nT  & $7\times{10^3\,{\rm km}}$ & $7\times{10^4\,{\rm km}}$%
& $7\times{10^5\,{\rm km}}$ & $4\times{10^6\,{\rm km}}$ \\
 & & & (0.005\,au) & (0.027\,au) \\
20\,nT & & & & $1\times{10^6\,{\rm km}}$ \\
50\,nT & 700\,km & 7,000\,km & $7\times{10^4\,{\rm km}}$%
& $4\times{10^5\,{\rm km}}$ \\\hline
decay probability & 0.0061 & 0.0250 & 0.0153 & 0.0047 \\
in 0.066\,au & & & & \\\hline
decay upto 0.934au & 0.9732 & 0.6731 & 0.2541 & 0.0698 \\
      up to 1au   & 0.9793  & 0.6980 & 0.2694 & 0.0745 \\\hline
\end{tabular}
\end{table}

\noindent
{\it Supplementary Information 4}

\begin{figure}[ht]
\centering
\includegraphics[width=110mm]{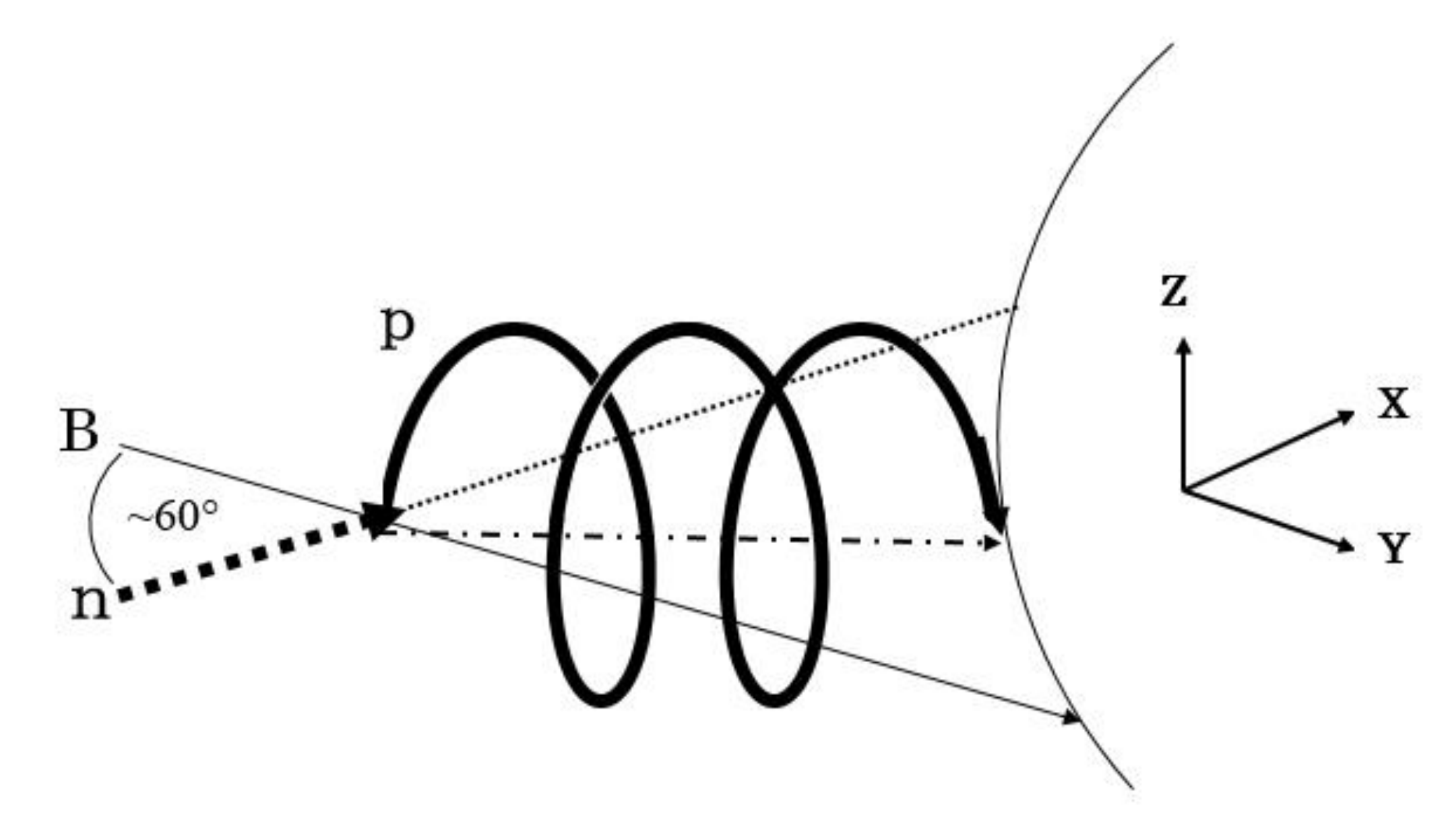}
\caption{%
The proton motion after neutron decay is pictorially shown.
Neutrons that are coming from the solar direction (the dashed line)
spontaneously decay into protons.
Then these protons receive the magnetic force by the IMF.
At that time the intensity of IMF was estimated as 20\,nT.
So protons with $Ep={\rm 6\,GeV}$ start the gyration motion
with a radius of $1\times{10^6}$\,km.
The equivalent $p_{\perp}$ is about 5\,GeV/$c$.
After a few rotations,
these protons arrive at the limit of the magnetosphere.
The IMF direction ($B$) and the GSE coordinate
are also shown in the figure.}
\label{sf6}
\end{figure}

Now let us examine the motion of the SNDPs.
Neutrons cross the interplanetary space almost straightforwardly
between the Sun and the Earth.
However as soon as they decay in the distance of $\ell$,
the protons of the decay product
immediately receive the electro-magnetic force,
so that the direction of the motion will be largely deflected
from the initial direction.
At that time the IMF direction crossed to $X$-axis
with $56^{\circ}$ to $60^{\circ}$.
Therefore, those protons received the equivalent transverse momentum
as 5\,GeV/$c$.
The rotation radius is estimated
as about $1\times 10^{6}$\,km or 0.0067\,au.
By this reason, before arrival at the magnetopause
(the head of the magnetosphere),
the SNDPs of 6\,GeV may rotate a few times before their arrival,
depending on the decay position.
A schematic view is shown in Figure~\ref{sf6}.

\noindent
{\it Supplementary Information 5}

\begin{figure}[ht]
\centering
\includegraphics[width=110mm]{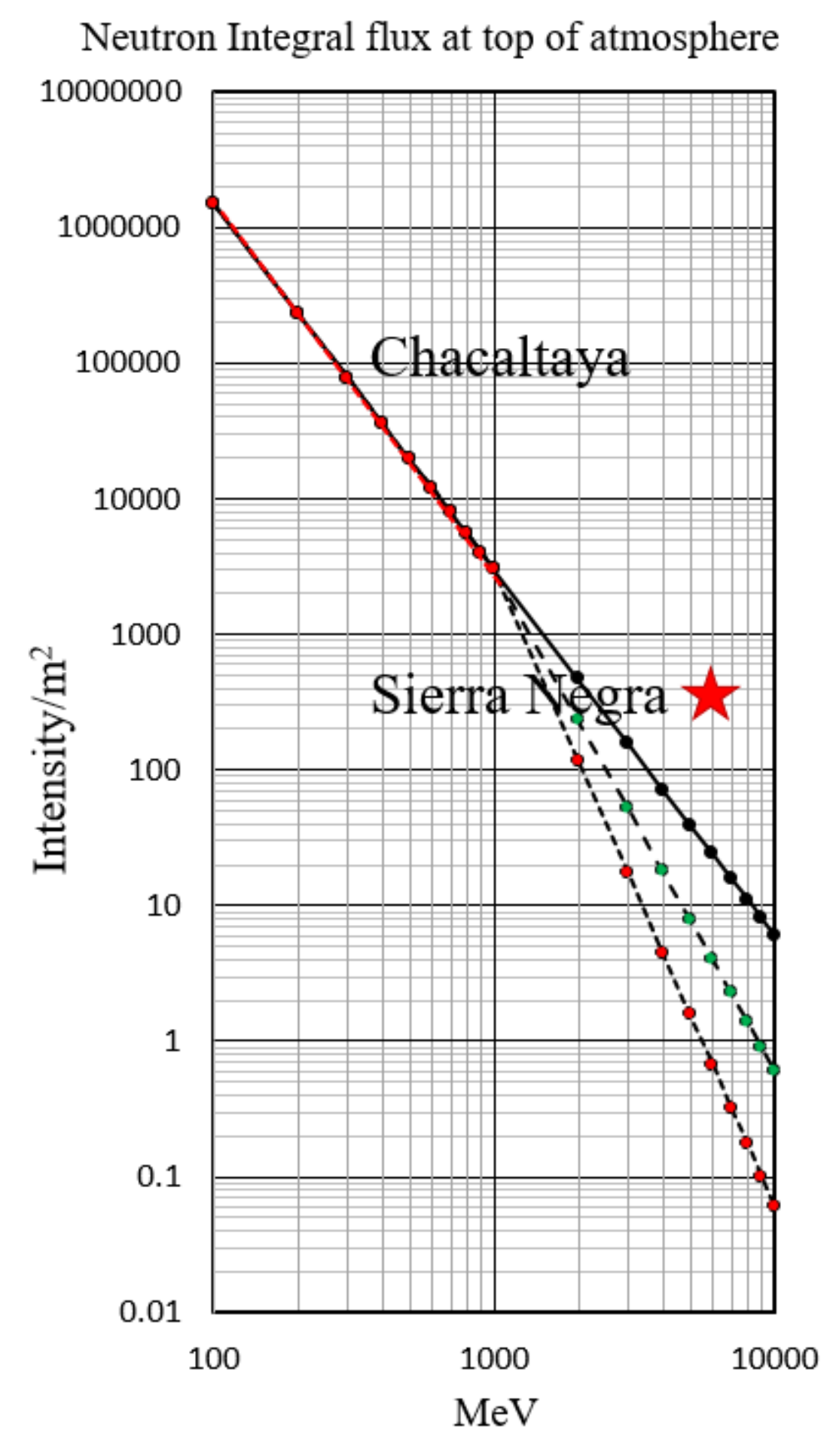}
\caption{%
The neutron production spectrum is shown
for the cases of three integral power index of $\gamma$.
The simple extended line shows
for the case of the power spectrum $\gamma=-2.7$,
and dashed line corresponds to the bending case
of the power index at 1\,GeV from $\gamma=-2.7$ to -3.7.
The dotted line represents the case
when the spectrum can be expressed by $\gamma=-4.7$
in the energy range between 1\,GeV and 10\,GeV (quite soft spectrum).
The estimated intensity at the space
integrated from the counting rate of Mt.~Sierra~Negra
is also shown by the star.
However the point represents the one-minute value of the integrated flux.
It corresponds to ($360 \pm 125\,{\rm neutrons/m^2\cdot min.}$).}
\label{sf7}
\end{figure}

In Figure~\ref{sf7},
we present the solar neutron spectrum measured at Mt. Chacaltaya.
The extended flux beyond 1\,GeV are also given
for the three cases of the spectrum
and we estimate the flux at 6\,GeV for each case.
The total intensity of 6\,GeV neutrons
estimated by the Mt.~Sierra~Negra detector is also plotted
on the line of $E_{n}=6$\,GeV.
The Chacaltaya neutron spectrum is presented by the integral spectrum
and the intensity was already converted into the flux
at the top of the atmosphere.

\nolinenumbers

\end{document}